\documentclass[twocolumn,showpacs,floatfix,prl]{revtex4}
\usepackage{graphicx}
\usepackage{dcolumn}
\usepackage{bm}
\usepackage{amsmath}
\usepackage{amssymb}
\usepackage{color}

\begin{document}

\title{Breached pairing in trapped three-color atomic Fermi gases}
\author{Beatriz Errea$^1$}
\author{Jorge Dukelsky$^1$}
\author{Gerardo Ortiz$^{2}$}
\affiliation{$^1$Instituto de Estructura de la Materia - CSIC, Serrano
123, 28006 Madrid, Spain \\
$^2$Department of Physics, Indiana University, Bloomington IN
47405, USA }

\begin{abstract}
We introduce an exactly solvable model for trapped three-color atom
gases. Applications to a cigar-shaped trapped cold fermions reveals a
complex structure of breached pairing phases. We find two competing
superfluid phases at weak and intermediate couplings, each one with two
color pair condensates, that can be distinguished from density profile
measurements.
\end{abstract}

\maketitle

Macroscopic coherent phenomena in matter, such as superconductivity and
superfluidity, are deep manifestations of wave mechanics with
consequences not only in technology but also for the fundamental
understanding of vacuum condensates in the standard model. Color
superconductivity is predicted to occur in quark matter at sufficiently
high density and low temperatures \cite{Quark1}. Quarks, having three
different colors (red, green, blue) and a strong attractive interaction,
allow for more diverse pairing patterns compared to the SU(2) Cooper
pairing in classic metallic superconductors. Such diversity, likewise,
makes it hard to establish the particular pairing symmetry favored by
nature. With the advent of ultracold trapped Fermi gases a window of
opportunities has opened to address some of these fundamental questions,
at least, in a qualitative fashion (see for example \cite{Giorg}). One
can certainly manipulate different atomic species and hyperfine states
to effectively generate multicolor Fermi gases with attractive
interactions.

It is the goal of this paper to investigate the superfluid
behavior of an {\it imbalanced} three-color Fermi gas by means of
an exactly-solvable pairing model of the Richardson-Gaudin (RG)
type, derived from the quadratic invariants of the SO(6) RG model
\cite{UshveAFS,Sie}. Previous studies using standard mean-field
\cite{Barions}, density matrix renormalization group \cite{dmrg},
or Bethe ansatz \cite{bethe} techniques concentrated on the
competition between a trionic or barionic phase and a color
superfluid phase. A main result of our work, from the standpoint
of collective behavior, is the competition between {\it
breached-pair}  (BP) and {\it unbreached-pair} (UP) superfluid
phases in a polarized multicolor Fermi gas. As in the two-color
case,  where density profiles have been recently investigated
experimentally \cite{Kett} and theoretically \cite{Theo}, an
analogue of the BP or Sarma phase \cite{Sarma,Wilc} appear in
multicolor polarized Fermi gases. In this case we find a complex
structure of breached pairing as well as the coexistence of two
pair condensates. While the possibility of coexistence of several
superfluid phases has been suggested in \cite{Coex} using an LDA
theory, we predict the existence two distinct {\it color fermionic
condensates}. Within our model this is a genuine effect, although
care must be exercised when contrasted to experiments since
interactions not included in our model could make this
 phase unstable against the formation of a fraction of bound trions in
the strong coupling limit. However, population imbalance as well
as the experimental realization of a stable three-color atomic gas
with different atomic masses and/or different Feshbach resonances,
as recently reported in experiments with balanced mixtures of
three components $^6$Li atoms \cite{Li1,Li2}, could stabilize it.

Consider the SU(3) color-symmetric Hamiltonian
\begin{equation}
H=\sum_{i}^{L}\varepsilon_{i}N_{i}-g\sum_{ii^{\prime
}}^{L}\sum_{\alpha }A_{i\alpha }^{\dagger} A^{\;}_{i^{\prime
}\alpha} \label{hamilton}
\end{equation}
for $L$ levels $i$ of energy $\varepsilon_{i}$, where
$\alpha$=(red($R$), green($G$), and blue($B$)) is the color index,
$N_{i}=\sum_\alpha N_{i\alpha}$ is  the number operator of the orbit
$i$, $A_{i\alpha}^{\dagger}=\sum_{\beta \gamma} \varepsilon_{\alpha
\beta \gamma }a_{i\beta}^{\dagger } a_{i\gamma}^{\dagger}$,
$A^{\;}_{i\alpha}=\left( A_{i\alpha}^{\dagger} \right)^{\dagger},$ are
the pair creators, and $g>0$ the pairing strength. Here
$a_{i\beta}^{\dagger}$ creates a (canonical) fermionic atom in level $i$
with color $\beta$, and  $\varepsilon_{\alpha \beta \gamma}$ is the
completely antisymmetric tensor in color space. The SO(6) algebra (15
generators) is completed by the $9$ particle-hole operators $C_{i,\alpha
\beta} \equiv a_{i\alpha}^{\dagger} a^{\;}_{i\beta}$. These 9 operators,
which include the number operators for the three different colors in the
level $i$, $N_{i\alpha}= C_{i,\alpha \alpha}$, close an U(3) subalgebra
of SO(6).

The Hamiltonian (\ref{hamilton}) has equal pairing strengths $g$ (or
scattering lengths), and equal single particle energies (masses) for the
three colors. It describes a three-color Fermi gas with attractive
contact interactions  in the low-density limit. Moreover, the SU(3)
symmetry is preserved and, thus, the eigenstates are organized in
degenerate SU(3) multiplets. The SU(3) symmetry, however, may be broken
by choosing a different combination of integrals of motion and, for
example, one can generate an integrable model of atoms with unequal
masses \cite{Note1}.

The exact solution of the SO(6) RG model, being an algebra of rank
three, depends on three sets of spectral parameters. The first set
includes the usual pair energies $e_{\alpha}$ of the SO(6) algebra,
while the other two sets, composed of the spectral parameters
$\omega_{\alpha}$ and $\gamma_{\alpha}$, are associated with the SU(3)
subalgebra of SO(6). The complete set of spectral parameters satisfy the
generalized Richardson equations
\begin{eqnarray}
\sum_{\beta (\not=\alpha )}^{M}\frac{2}{e_{\beta }-e_{\alpha
}}-\sum_{\beta }^{N_{B}}\frac{1}{\omega _{\beta }-e_{\alpha
}}+\sum_{i}^{L}\frac{\nu _{i}-1}{2\varepsilon _{i}-e_{\alpha }}
&=&-\frac{1}{4g}  \notag \\
\sum_{\beta (\not=\alpha )}^{N_{B}}\frac{2}{\omega _{\beta }-\omega
_{\alpha }} -\sum_{\beta }^{M}\frac{1}{e_{\beta }-\omega _{\alpha
}}-\sum_{\beta }^{Q}\frac{1}{\gamma _{\beta }-\omega_{\alpha }} &=&0
\label{Req} \\
\sum_{\beta }^{Q}\frac{2}{\gamma _{\beta }-\gamma _{\alpha }
}-\sum_{\beta (\not=\alpha )}^{N_{B}}\frac{1}{\omega _{\beta
}-\gamma _{\alpha }}-\sum_{i}^{L}\frac{\nu _{i}}{2\varepsilon
_{i}-\gamma _{\alpha }} &=&0.  \notag
\end{eqnarray}

The number of spectral parameters in each set is determined by the
number of particles of each color $N_{\alpha}$, and by the total
seniority quantum number $\nu$: $M=\left( N-\nu \right) /2$ and
$Q=(N_{B}+N_{R}-N_{G}+\nu )/2$, where we have assumed, without loss of
generality,  $N_G \geq N_R \geq N_B$. The seniority of level $i$, $\nu
_{i}$, counts the number of unpaired fermions, and is defined from
$A_{i\alpha }\left\vert \nu _{i}\right\rangle =0$, $N_{i}\left\vert
\nu_{i} \right\rangle =\nu _{i}\left\vert  \nu _{i}\right\rangle$,
$\nu_{i}=0,1$. The total seniority is $\nu =\sum_{i}\nu _{i}$.
Hamiltonian (\ref{hamilton}) preserves the seniority since it can create
or destroy pairs of particles conserving the number parity of the level,
i.e., for a given configuration each level has an even (odd) number of
particles where $\nu_{i}=$0 (1).

The eigenvalues of Hamiltonian (\ref{hamilton}) are
\begin{equation}
E=\sum_{\alpha }^{M}e_{\alpha }+\sum_{i=1}^{L}\varepsilon _{i}\nu
_{i},
\end{equation}
and only depend on the parameters $e_{\alpha}$. The corresponding
eigenfunctions, though, are determined by the three sets of parameters.
The solutions of the Richardson equations defines a basis which spans
the complete many-body Hilbert space of the system.

Assume a polarized gas composed of $N=N_G+N_R+N_B$ fermionic atoms. In
the weak coupling limit the energy levels are filled up to the Fermi
energy for each color $\varepsilon_{N_{\alpha}}$, a situation depicted
in Fig. $1$ {\bf A} for $N_G=80$, $N_R=50$ and $N_B=20$. In this case
the seniorities $\nu_i$ are equal to 1 for $i\leq 20$ and $50 < i \leq
80$, and they are $0$ for $20 < i \leq 50$ and $i > 80$, defining a
clear separation of the Hilbert space into regions of odd particle
states ($\nu_i=1$) and regions of even particle states ($\nu_i=0$). When
the pairing interaction $g$ is switched on, $R$-$B$ pairs from the first
region ($i \leq 20$) cannot scatter to the second region ($20 < i \leq
50$) due to Pauli blocking, and they have to jump this forbidden region
to scatter into the third region $50 < i \leq 80$. Analogously, $G$-$R$
pairs of the second region have to jump the third forbidden region to
scatter into the fourth region ($i>80$). This configuration, that we
call BP state, turns out to be the ground state (GS) at weak coupling.
For larger values of $g$ other configurations compete with the BP state.
Those configurations, at the cost of increasing their kinetic energy,
reduce the effect of Pauli blocking, therefore, facilitating the pair
scattering into interior level regions. In particular, the UP state
depicted in panel {\bf B} has no blocked interior region, and  will be
the GS of the system at strong coupling.
\begin{figure}[htb]
\vspace*{-.5cm} 
\includegraphics[angle=0,width=9.0cm,scale=1.0]{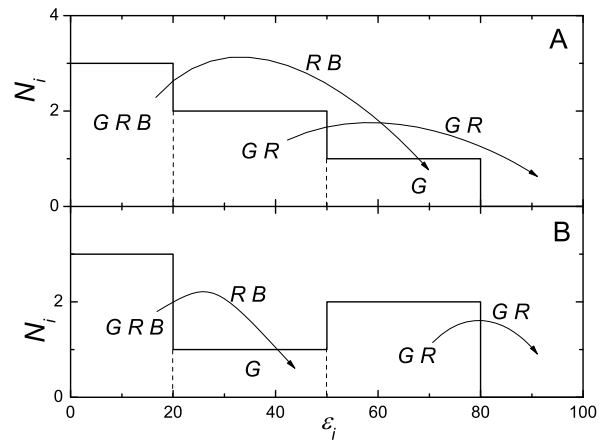}
\caption{Occupation number for the BP state (A) and the UP
state (B). }
\label{fig1}
\end{figure}

\begin{figure}[htb]
\vspace*{-1.0cm} 
\includegraphics[angle=0,width=9.0cm,scale=1.0]{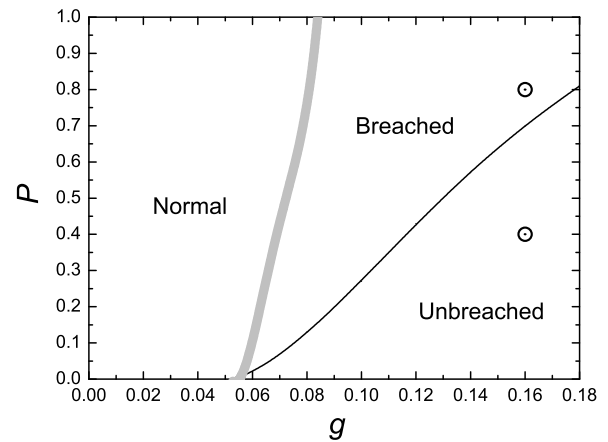}
\caption{Phase diagram of a 1D partially polarized trapped Fermi gas
with $N=150$, $N_R=50$, $ P=(N_G - N_B)/(N_G + N_B)$, and $L=500$. }
\label{fig2}
\end{figure}
Although Hamiltonian (\ref{hamilton}) is exactly solvable in any
dimension, for simplicity we will consider a system of $N=150$
($N_R=50$) fermionic atoms trapped by a 1D harmonic potential (of
frequency $\omega$) with an energy cutoff at $E_{\sf cut}=500 \hbar
\omega$, implying $L=500$ threefold degenerate single particle levels.
In the weak coupling limit, the Richardson equations (\ref{Req})
decouple into independent sets of equations, each one related to the
single particle level, partially or fully occupied, as discussed above.
These equations can be solved analytically producing three sets of
spectral parameters. These parameters are used as an initial seed in an
iterative procedure in which the coupling constant $g$ is systematically
increased by using the solution of the previous step as the new initial
guess. In this way the initial solution at weak coupling is evolved up
to the desire value of $g$. We performed extensive calculations to
determine the quantum phase diagram of this system as a function of the
pairing strength $g$ and the polarization $P=(N_G - N_B)/(N_G + N_B)$.

Two color superfluid phases emerge as a function of color asymmetry and
pairing strength (see fig. \ref{fig2}). A first order quantum phase
transition, due to level crossing, separates the BP and UP superfluid
phases, which are labeled by different sets of seniority quantum
numbers. On the other hand, there is a smooth crossover between the two
superfluid phases and a normal, fluctuation-dominated state depicted by
a thick grey line \cite{Note2}. The normal Fermi-liquid-like state is
dominated by pairing fluctuations which are fully taken into account by
the exact solution. We adopted the criterium that the normal region
extends, for a given $P$, from $g=0$ up to the value of $g$ for which
$10 {\%}$ of the pair energies $e_\alpha$ are complex, meaning that the
condensate fraction $f$ is $\leq 0.1$ \cite{ours1}. To study the
correlations and structure of the BP and the UP states we have chosen
two particular points, marked with open circles in fig. \ref{fig2}.
Notice that within our model seniority is a conserved quantum number,
disfavoring the formation of bound trionic molecular states which are
more likely to appear for very low population imbalance and very strong
coupling.
\begin{figure}[htb]
\vspace*{-.5cm} 
\includegraphics[angle=0,width=9.0cm,scale=1.0]{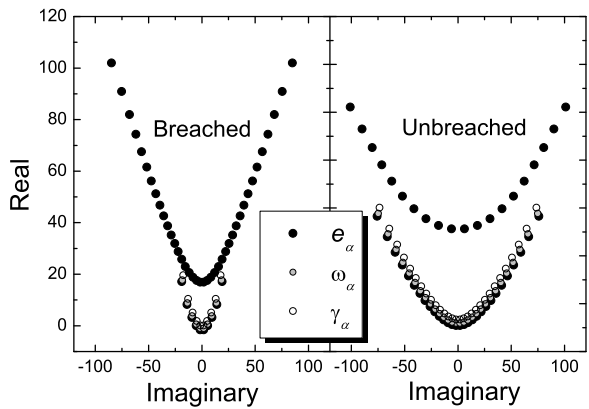}
\caption{Spectral parameters of the BP state with $P=0.8$ and the
UP state with $P=0.4$, for $g=0.16$. }
\label{fig3}
\end{figure}

Figure \ref{fig3} shows the three sets of spectral parameters for these
two states. Note that for this value of $g=0.16$ all pair energies are
complex with a positive real part, implying that the condensate
fraction  is $f=1$, and all pairs behave as Cooper resonances as opposed
to bound molecules of a Bose Einstein condensate (BEC) \cite{ours1}. In
both cases the pair energies form two separate arcs in the complex
plane, indicating the existence of {\it two color fermionic pair
condensates}. In the strong coupling limit the two arcs coalesce into a
single arc corresponding to a single BEC condensate with negative real
part of their pair energies. The lower arc is overlapping with arcs of
the two other spectral parameters $\omega_\alpha$ and $\gamma_\alpha$
which account for the couplings in the SU(3) color subspace. The
interpretation is that the upper arc of isolated pair energies
$e_\alpha$ describes $G$-$R$ Cooper pairs, while the lower arc
corresponds to $R$-$B$ Cooper pairs. An analysis in terms of the
eigenvalues of the two-body density matrix would lead to a macroscopic
eigenvalue of the $G$-$R$ and the $R$-$B$ pair density matrices and no
macroscopic eigenvalue in the $G$-$B$ pair density matrix. The
appearance of these two condensates will be reflected in the occupation
probabilities that can be calculated using the Hellmann-Feynman theorem
on the integrals of motion as will be explained in a forthcoming paper
\cite{Errea}.
\begin{figure}[htb]
\vspace*{-.5cm} \hspace*{-0.8cm}
\includegraphics[angle=0,width=10.0cm,scale=1.0]{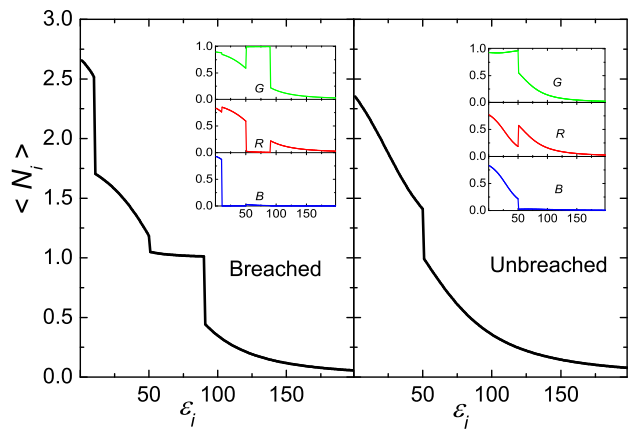}
\vspace*{-.5cm}
 \caption{(Color online) Occupation numbers of the BP state with $P=0.8$ and the UP state with
$P=0.4$, for $g=0.16$. In the inset we display the different color contributions to $\langle N_i \rangle$.}
\label{fig4}
\end{figure}

The occupation probabilities, $\langle N_i \rangle$, for both states are
depicted in fig. \ref{fig4}. In the BP state panel we see how the
$G$-$R$ pairs avoid the region $50 < i \leq 90$, Pauli blocked by the
$G$ atoms, to scatter off into the region $i > 90$. Analogously, though
less evident due to the smaller number of pairs, the 10 $R$-$B$ pairs
avoid the region $10 < i \leq 50$, blocked by the $R$ atoms, to scatter
off into the region $i > 50$. We may realize that in the latter case the
blocking is not perfect, unlike in the two-component gases, because of
the depletion of the $R$ atoms in the $10 < i \leq 50$ region due to
$G$-$R$ pairing. The $G$-$B$ pairing is prevented from being realized
due to two consecutive blocked regions. On the contrary, in the UP state
pairs do not have to avoid blocked regions. The $G$-$R$ pairs in the $50
< i \leq 70$ region scatter off into the regions $i > 70$, and $R$-$B$
pairs in the region $i \leq 30$ scatter off into the region $30 < i \leq
50$ as well as into the depleted region $i > 50$. These two different
physical scenarios manifest in the arc geometry of the spectral
parameters in fig. \ref{fig3}. While the BP state has an inner
condensate of 10 $R$-$B$ and an outer condensate of 40 $G$-$R$ pairs,
the UP state has 30 condensed $R$-$B$ and 20 condensed $G$-$R$ pairs.
\begin{figure}[htb]
\vspace*{-.5cm} 
\includegraphics[angle=0,width=9.0cm,scale=1.0]{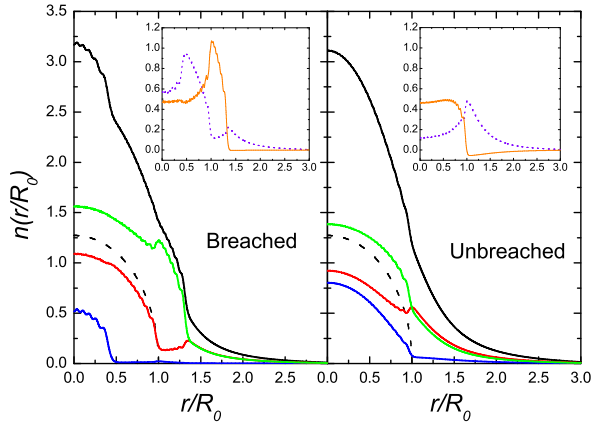}
\caption{(Color online) Radial density profiles for the BP and UP states
for $g=0.16$. The dash curve corresponds to the Thomas-Fermi
approximation for the $R$ atoms. The insets show the density differences
between the $R$ and $B$ (dash dark gray), and $G$ and $R$ (solid light
gray) species.}
\label{fig5}
\end{figure}

An experimental way to uncover the nature of the color superfluid
correlations consists in measuring the density clouds of the trapped
Fermi gas. Figure \ref{fig5} shows the radial density profiles for these
two states normalized to the $R$ density as a function of the distance
$r$ from the center of the trap, in units of the Thomas-Fermi radius
$R_0$ of the $R$ species. The insets display the density differences
between the $R$ and $B$ (dash dark gray), and $G$ and $R$ (solid light
gray) species. In both cases the outer region is dominated by $G$-$R$
pairing that enforces a maximum overlap of the $G$ and $R$ components of
the wavefunction. However, the BP state shows a richer structure in the
inner region. In particular, a clear signal of a BP state appears as a
peak in the density profile of the majority atoms $G$, which is
magnified in the $G$ and $R$ density difference. Two peaks appear in the
$R$ and $B$ density difference consistent with the decay and revival of
the $B$ density in regions of $R$-$B$ pairing. The structure of the
density differences is smoother in the UP phase due to the absence of
breached pairing. Regions of constant energy differences signal the
dominance of the corresponding pairing phase. In the UP state for $0.0 <
r/R_0 < 0.5$ it is dominated by a $R$-$B$ superfluid, while for
$r/R_0>1.0$ we have a $G$-$R$ superfluid.

One can think about measuring the fraction of each of the two color pair
condensates by exploiting the differences in the color-pair dependent
Feshbach resonances \cite{Li2}. It would then be possible to use the
ramp technique as described in \cite{ramp}, sweeping the magnetic field
such that only one class of pairs are transformed into bound molecules,
allowing for the determination of the corresponding fraction of the
condensate. A bigger ramp would then transform all pairs into molecules,
therefore, allowing measurement of the complete fraction of the
condensate.

We acknowledge fruitful discussions with G. G. Dussel. This work was
supported in part by the grant FIS2006-12783-C03-01 of the Spanish DGI.
B.E. was supported by the Spanish CE-CAM.

\end{document}